\documentclass[preprint,12pt,authoryear]{elsarticle}

\usepackage{amssymb}
\usepackage{amsmath}

\usepackage{array}
\usepackage{changepage}
\usepackage{float}
\usepackage{hyperref}
\usepackage[colorinlistoftodos]{todonotes}
\usepackage{xcolor}

\journal{Computers \& Security}

\begin{document}

\begin{frontmatter}



\title{Operationalizing Cybersecurity Knowledge: Design, Implementation \& Evaluation of a Knowledge Management System for CACAO Playbooks\tnoteref{t1}} 

\tnotetext[t1]{This preprint has not been peer-reviewed. It is a preliminary version of a research article that has been submitted for journal publication. The final, peer-reviewed version may differ from this preprint.}

 \author [label1]{Orestis Tsirakis}
 \author [label2,label3]{Konstantinos Fysarakis}
 \author [label3]{Vasileios Mavroeidis}
 \author [label1]{Ioannis Papaefstathiou}
 
 \affiliation[label1]{organization={School of Electrical and Computer Engineering, Aristotle University of Thessaloniki},
             city={Thessaloniki},
             country={Greece}}
 \affiliation[label2]{organization={Automaton Technologies Ltd},
             city={Nicosia},
             country={Cyprus}}

\affiliation[label3]{organization={University of Oslo},
             city={Oslo},
             country={Norway}}








\begin{abstract}
Modern cybersecurity threats are growing in complexity, targeting increasingly intricate and interconnected systems. To effectively defend against these evolving threats, security teams utilize automation and orchestration to enhance response efficiency and consistency. In that sense, cybersecurity playbooks and workflows are key enablers, providing a structured, reusable, and continuously improving approach to incident response, enabling organizations to codify requirements, domain expertise, and best practices and automate decision-making processes to the extent possible. The emerging Collaborative Automated Course of Action Operations (CACAO) standard defines a common machine-processable schema for cybersecurity playbooks, facilitating interoperability for their exchange and ensuring the ability to orchestrate and automate cybersecurity operations. However, despite its potential and the fact that it is a relatively new standardization work, there is a lack of tools to support its adoption and, in particular, the management and lifecycle development of CACAO playbooks, limiting their practical deployment. Motivated by the above, this work presents the design, development, and evaluation of a Knowledge Management System (KMS) for managing CACAO cybersecurity playbooks throughout their lifecycle. It provides essential tools to improve cybersecurity maturity, strengthens collaboration and coordination within and across organizations, and streamlines playbook management. Using open technologies and standards, the proposed approach fosters standards-based interoperability and enhances the usability of state-of-the-art cybersecurity orchestration and automation primitives. To encourage adoption, the resulting implementation is released as open-source, which, to the extent of our knowledge, comprises the first publicly available and documented work in this domain, supporting the broader uptake of CACAO playbooks and promoting the widespread use of interoperable automation and orchestration mechanisms in cybersecurity operations.
\end{abstract}




\end{frontmatter}



\section{Introduction}
\label{sec1}

The rapid advancement of technology in recent decades has led to a sharp rise in various cybersecurity incidents, including unauthorized access, denial-of-service (DoS) attacks, malware infections, data breaches, and phishing. These threats pose substantial financial risks to individuals, businesses, and organizations. Consequently, implementing and enforcing a robust cybersecurity strategy is essential for organizations to effectively mitigate these risks (\citet{ahsan2022cybersecurity}).

In this context, automation plays a crucial role in cybersecurity by enhancing incident response capabilities. Automating security responses reduces reaction times and improves accuracy, enabling organizations to address threats swiftly and efficiently. A key automation technology in this domain is the use of playbooks, which define standardized procedures for handling security incidents and guide organizations through predefined response actions (\citet{schlette2024you}).

Sharing playbooks across organizations has the potential to significantly enhance cybersecurity resilience, particularly for those with limited resources, expertise, or experience in managing cyber incidents (\citet{akbari2022sasp}). However, traditional playbooks are often tailored to specific organizations and lack interoperability, making them neither machine-readable nor easily shareable. Standards such as the Collaborative Automated Course of Action Operations (CACAO) (\citet{CACAO-Security-Playbooks-v2.0}) help address this issue by providing a structured framework for the creation and management of playbooks. Despite these advancements, effectively managing playbooks (CACAO playbooks or others) throughout their lifecycle—including creation, retrieval, execution, and sharing—remains a significant challenge.

Building on the above, this paper focuses on designing, developing, and evaluating a Knowledge Management System (KMS) tailored for the efficient storage, retrieval, and management of CACAO playbooks. The proposed system aims to enhance the use of playbooks, promote their broader adoption, and enable automation—ultimately improving effectiveness and efficiency in cybersecurity operations, including incident response. The approach is derived from a thorough analysis of existing standards, technologies, and knowledge management models, which are adapted to meet the specific requirements of playbook-centric knowledge management. Furthermore, the implementation of the Playbook KMS is released as open-source to encourage adoption within the cybersecurity community and facilitate greater interoperability and collaboration among cybersecurity teams.

The remainder of this paper is structured as follows: Section 2 provides background information on automation, security playbooks, and knowledge management. Section 3 outlines the design of the proposed knowledge management system. Section 4 presents the proof-of-concept implementation, detailing its core components and system evaluation. Section 5 discusses key limitations and offers a set of propositions. Finally, Section 6 concludes the paper with closing remarks.

\section{Background \& Motives}
\label{sec2}

\subsection{Cybersecurity Automation - Potential Solutions \& Challenges}

Automation plays a pivotal role in enhancing organizational security by enabling a proactive approach to identifying, analyzing, and mitigating potential risks efficiently and effectively. In the realm of cybersecurity, the emergence of advanced and sophisticated attacks has become an escalating concern for companies and organizations. Traditional, manual approaches to incident response are increasingly proving insufficient, reinforcing the necessity for automated operations to counter the rising threat landscape.

One of the primary advantages of automation is its ability to enhance the efficiency of security operations, thereby increasing productivity and minimizing the time required to resolve security issues. By automating repetitive and time-intensive tasks, security analysts can dedicate their efforts to more complex activities, such as threat analysis and research, contributing to organizational knowledge and value. Moreover, automation enables organizations to identify and respond to potential security incidents with greater accuracy, effectively mitigating risks and reducing the likelihood of human error (\citet{mohammad2018security}).

\subsubsection{Security Orchestration, Automation, and Response}

Security Orchestration, Automation, and Response (SOAR) encompasses technologies and software that empower organizations to gather data on security threats from diverse sources and automate incident response processes, reducing the need for human intervention. SOAR solutions are specifically designed to enhance the efficiency and effectiveness of security operations by streamlining response capabilities and offering tools to manage security operations across an organization's digital environment. These solutions facilitate the creation and execution of playbooks, automate routine tasks, and coordinate the actions of various security tools, thereby enabling the management of a higher volume of incidents with improved speed and accuracy (\citet{akbari2022sasp, kraeva2021application}).

There is a proliferation of commercial solutions in the SOAR market that rely on different playbook types, such as Chronicle SOAR, Cyware SOAR, IBM QRadar SOAR and Fortinet FortiSOAR, and some other open-source systems such as SOARCA (\citet{SOARCA-github}), which uses the CACAO standard, and Shuffle Automation (\citet{shuffler-io}).

\subsubsection{Playbooks \& the CACAO Standard}

Incident response playbooks have become indispensable in the standardization and automation of cybersecurity operations, offering structured frameworks to systematically guide incident response efforts. Also known as Course of Action (CoA) playbooks, they comprise predefined procedures and step-by-step actions designed to address a broad spectrum of security incidents, from routine breaches to advanced cyberattacks (\citet{mavroeidis2021integration}). These playbooks integrate the collective expertise, experience, and best practices of cybersecurity professionals, ensuring a structured and effective approach to incident detection, analysis, and mitigation. A key challenge lies in the continuous need to update and adapt playbooks to keep pace with the ever-evolving cyber threat landscape, organizational process changes, and technological advancements.

In the modern cybersecurity environment, where threats are pervasive and rapid response is crucial, the significance of cybersecurity playbooks cannot be overstated. These playbooks not only provide a structured framework for automated incident response but also foster knowledge sharing, collaboration, and innovation within cybersecurity teams. As cyber threats continue to evolve, playbooks will remain fundamental in strengthening organizational defense and mitigating security incidents effectively (\citet{fysarakis2023phoeni2x}).

Recent studies on cybersecurity highlight the importance of incident response playbooks but also reveal significant gaps in their design, usability, and practical application. One study (\citet{schlette2024you}) evaluated 1217 playbooks and conducted surveys and interviews, uncovering ambiguities in how playbooks are defined and adapted across organizations. Another study (\citet{stevens2022ready}) examined standards like Integrated Adaptive Cyber Defense (IACD) and NIST Computer Security Incident Handling Guide, finding that while these playbooks are valuable, they often lack detailed, practical guidance for less experienced technicians, creating a gap between theoretical frameworks and real-world applicability. Both studies emphasize the need for more detailed, user-friendly playbooks with step-by-step instructions, highlighting the importance of adaptability to meet organizational-specific needs. Furthermore, research on standardizing incident response frameworks (\citet{schlette2021comparative}) underscores the importance of integrating structured data formats and aligning playbooks with incident response standards. It introduces key concepts for organizations to evaluate and select standards that enhance decision-making and improve response effectiveness. Collectively, these studies stress the critical role of usability, adaptability, and standardization in developing playbooks that bridge theoretical robustness with practical usability, ultimately supporting effective incident response strategies in diverse organizational contexts.

In recent years, several open standards have been developed to support the automation of incident response and the sharing of playbooks (\citet{empl2024generating}). Open standards are publicly available specifications designed to promote interoperability and seamless data exchange between diverse products and services. These standards are established, adopted, and maintained through a collaborative, consensus-driven process, ensuring broad applicability and integration across various cybersecurity frameworks (\citet{fysarakis2022blueprint}).

The introduction of CACAO, developed under the auspices of the Organization for the Advancement of Structured Information Standards (OASIS), has provided security practitioners with a standardized framework to enhance the actionability of Cyber Threat Intelligence (CTI) through the creation and sharing of both machine- and human-readable playbooks. Before CACAO, playbook development relied on proprietary technologies, which hindered their exchange and limited users' ability to understand and adapt them to their specific needs (\citet{mavroeidis2021integration}). By offering a standardized specification schema, CACAO enables the structured and consistent creation and dissemination of cybersecurity playbooks, ensuring interoperability across organizational and technological boundaries (\citet{CACAO-Security-Playbooks-v2.0}).

CACAO is designed with flexibility in mind, allowing for diverse use cases and seamless integration of automated actions. Users can develop playbooks with varying levels of abstraction, detail, and complexity, tailoring them to their specific objectives, operational requirements, and expertise. These playbooks facilitate the coordination of multiple operational roles, defense processes, and entities involved in their execution (\citet{zych2023reviewing}). Their applicability extends across a broad spectrum of cybersecurity activities, including threat hunting, detection, investigation, prevention, mitigation, remediation, and simulated attacks. Furthermore, execution can be triggered by automated or manual events, human observation, or predefined schedules, such as recurring or periodic tasks (\citet{mavroeidis2021integration}).

\begin{figure}[h]
\centering
\includegraphics[scale=0.5]{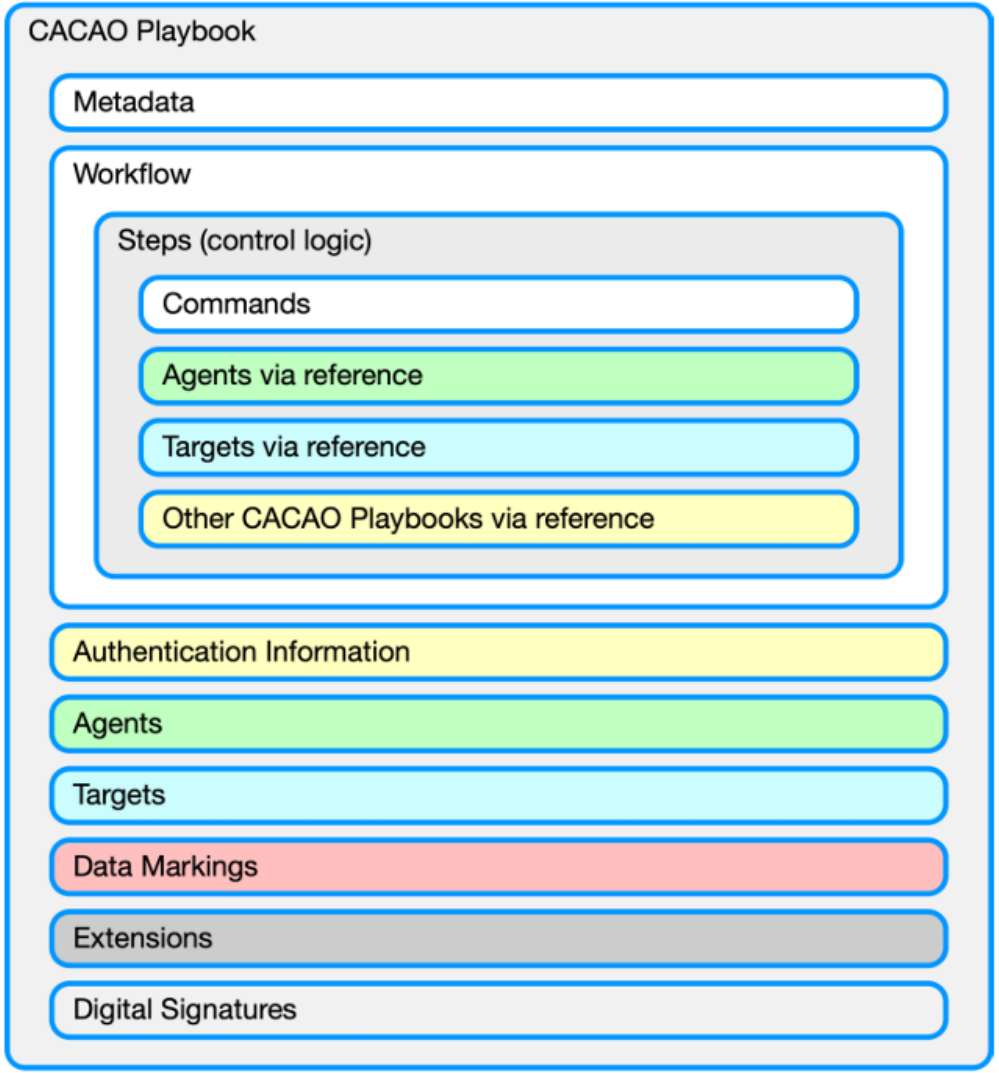}
\caption{CACAO Structure}\label{cacao_structure}
\end{figure}

CACAO draws inspiration from the Business Process Model and Notation (BPMN) design model (\citet{omg2011bpmn}), adapting it to establish a standardized format in JavaScript Object Notation (JSON) specifically for cybersecurity applications. At a high level (Figure \ref{cacao_structure}), CACAO playbooks consist of several key components: metadata, workflow steps that define the logic for controlling command execution, a set of executable commands, targets responsible for processing and executing these commands, data markings that specify handling and sharing requirements, and extensions that enable the incorporation of additional functionality. Moreover, CACAO playbooks support digital signatures to ensure integrity and authenticity, allowing the signature to be embedded within the playbook or stored separately as an independent signature.

Recent research on the CACAO standard highlights its growing importance in promoting structured, shareable cybersecurity measures for dynamic threat response and mitigation. One study, which comprises the first documented work on CACAO playbooks, emphasizes the integration of CACAO into broader frameworks to enhance the transformation of ambiguous CTI into actionable, automated processes, fostering interoperability across organizations (\citet{mavroeidis2021integration}). In addition, the authors provide practical implementations in integrating CACAO with other standards and open source solutions like Structured Threat Information eXpression (STIX), Malware Information Sharing Platform (MISP), and a formal ontology defined in Web Ontology Language (OWL) for STIX version 2.1, subsumed by the Threat Actor Context (TAC) ontology. Research by (\citet{zych2023reviewing}) investigates the advantages, feasibility, and complexity of using BPMN to graphically represent the complex workflows within CACAO playbooks to improve clarity and operational efficiency. Similarly, a study by (\citet{settanni2023model}) proposes a model leveraging CACAO templates to automate threat remediation, streamlining response actions and enabling efficient sharing among stakeholders. The development of tools like the Semantic web-based Approach for management of Sharable cybersecurity Playbooks (SASP) further advances CACAO applications, significantly enhancing playbook lifecycle management (\citet{akbari2022sasp}). Additionally, research from (\citet{empl2024generating}) expands the utility of CACAO by adapting it for industrial control systems (ICS), showcasing the flexibility of the standard to meet specific industry needs. These advancements underline CACAO's critical role in automating cybersecurity operations, ensuring timely and effective responses while supporting collaborative defense strategies across diverse organizational and industrial contexts.

Efforts to automate incident response procedures through SOAR systems and open standards underscore the vital role of playbooks in minimizing damage and facilitating the recovery of compromised systems (\citet{kraeva2021application}). However, fully orchestrating and automating the entire lifecycle of cybersecurity incident response playbooks—including preparation, detection, and post-incident reporting—remains a subject that warrants further research and development and remains use-case-based (\citet{lekidis2024towards}).

\subsubsection{Automation \& Playbook Adoption Challenges}

As highlighted, the effective management and dissemination of knowledge are critical to strengthening organizational defenses against cyber threats. The reliance on proprietary technologies has constrained the development and sharing of security playbooks, limiting their widespread adoption. The introduction of open standards such as CACAO is transforming the structuring and exchange of playbooks, enabling cybersecurity professionals to enhance their responsiveness to threats and significantly improve their defensive capabilities.

Despite the promising advancements enabled by standards such as CACAO, significant challenges remain in ensuring the quality and usability of shared playbooks (\citet{settanni2023model}). Issues such as ambiguities in version management, outdated information, and design inconsistencies present substantial obstacles to their effective deployment in cybersecurity operations. The absence of advanced knowledge management tools further exacerbates these challenges, highlighting the need for robust solutions capable of efficiently storing, organizing, and distinguishing playbooks from other incidental information. (\citet{mitre-cybersecurity-ops-center}).

This paper (following the scientific method) aims to address the existing gap in the storage, retrieval, and comprehensive management of CACAO playbooks throughout their lifecycle by designing and delivering an efficient and user-friendly KMS. By leveraging modern standards and advanced data management techniques, this work seeks to equip organizations with the necessary tools to enhance collaboration and strengthen their defenses against cyber threats. The practical objective is to develop a robust platform for storing and managing CACAO playbooks, thereby facilitating their accessibility, usability, and seamless sharing.

\subsection{Classifying CACAO Metadata in the context of the KMS}

In our KMS we have efficiently utilized the CACAO metadata to improve the organization, retrieval and overall management of playbooks. The CACAO standard defines metadata as a crucial component, enabling efficient indexing and searchability of playbooks. CACAO metadata can be split into different categories, regarding their function and the information they provide.

\begin{enumerate}
  \item Administrative Metadata: Provide essential information for managing playbooks throughout their lifecycle, aiding to identification, version control, access management and sharing. 
  
  \underline{Properties:} id, type, spec\_version, created\_by, created, modified, markings and revoked.
  \item Descriptive Metadata: Enhance searchability by providing human-readable details about the purpose and context of the playbook.
  
  \underline{Properties:} name, description, external\_references and labels.
  \item Technical Metadata: Provide detailed technical information about how the playbooks are structured and designed to function. 
  
  \underline{Properties:} playbook\_processing\_summary, playbook\_types and playbook\_activities.
  \item Operational Metadata: Provide insights into the practical application of playbooks in real-world scenarios. 
  
  \underline{Properties:} priority, severity, impact and  industry\_sectors.
\end{enumerate}

This categorization of metadata ensures a structured approach to managing CACAO playbooks effectively throughout their lifecycle. Administrative Metadata provide fundamental details for identification, version control, access management, and sharing, enabling seamless governance and collaboration. Descriptive Metadata enhance the discoverability of playbooks by offering human-readable information about their purpose and context. Technical Metadata define the structural and functional aspects of playbooks, detailing their processing, types, and activities. Lastly, Operational Metadata offer insights into the real-world applicability of playbooks, outlining their priority, severity, impact, and relevance to specific industry sectors. Together, these metadata categories support efficient playbook management, execution, and interoperability.

\subsection{Knowledge Management Systems}

Knowledge Management (KM) is a systematic process aimed at collecting, sharing, and effectively utilizing knowledge within an organization. Its primary objective is to bridge the gap between knowledge holders and those seeking information, ultimately enhancing organizational performance and fostering innovation. This is achieved not only by facilitating individual skill acquisition but also by ensuring that knowledge is accessible to the entire workforce. The core responsibilities of KM include knowledge storage, improving accessibility, and strengthening the infrastructure for knowledge management and dissemination. Given its critical role in organizational efficiency, KM is essential for sustaining growth, adaptability, and informed decision-making (\citet{smartsheet-knowledge-management, mit-ecommerce}).

\begin{enumerate}
  \item It increases idea generation and collaboration: KM promotes an environment where new ideas can be generated and shared, leading to innovation and continuous improvement.
  \item It increases efficiency: With a well-implemented KM system, members of any organization spend less time searching for information and more time applying it, thus enhancing overall productivity.
  \item It keeps existing knowledge secure: Organization is a very important aspect of KM; otherwise, the stored knowledge will be impossible to identify and use and will likely be lost over time.
  \item It improves decision-making: By ensuring that the right knowledge is accessible to the right people at the right time, organizations can improve their decision-making processes.
\end{enumerate}

The concept of KM has evolved significantly over the past decades. Initially centered on basic data management, its scope expanded as organizations recognized the strategic value of knowledge in driving efficiency and innovation. KM models offer a structured framework for managing knowledge throughout its lifecycle, outlining the stages from creation to utilization to ensure that critical information is effectively captured and leveraged (\citet{smartsheet-knowledge-management}). Given their systematic approach, KM models hold significant potential for optimizing the management of CACAO playbooks across their entire lifecycle. Several models have been proposed (Table \ref{km_models}), each of which emphasizes different aspects of the KM process. A comparison of those models is presented in Table \ref{km_models_comparison}.

\begin{table}[h]
\centering
\begin{tabular}{|>{\centering\arraybackslash}p{3cm}|>{\centering\arraybackslash}p{3cm}|>{\centering\arraybackslash}p{3cm}|>{\centering\arraybackslash}p{3cm}|}
\hline
\textbf{Wigg (1993)} & \textbf{Zack (1996)} & \textbf{Bukowitz \& Williams (2000)} & \textbf{McElroy (2003)} \\
\hline
Creation & Acquisition & Get & Learning \\
\hline
Sourcing & Refinement & Use & Validation \\
\hline
Compilation & Store & Learn & Acquisition \\
\hline
Transformation & Distribution & Contribute & Integration \\
\hline
Application & Presentation & Assess & Completion \\
\hline
\end{tabular}
\caption{KM Lifecycle Models}\label{km_models}
\end{table}

\textbf{Wiig Model (1993):} This model is based on the idea that information needs to be organized in order to be useful. Thus, this model is primarily concerned with organizing all data once it is codified, but also describes how knowledge is created, saved, combined (with other stored knowledge), and then extended inside the organization. The phases of the Wiig model are creation, sourcing, compilation, transformation, and application.

The Wiig model has been utilized as a foundational framework for KM implementation projects to enhance organizational practices (\citet{naderi2024crisis}), guide chatbot development for dynamic web-based KM systems in small-scale agriculture (\citet{ong2021review}), and support the design of adaptable KM systems in the oil refining industry without significant limitations (\citet{ekaterina2021development}).

\textbf{Zack Model (1996):} The Zack model emphasizes a logical and standardized procedure when moving on to each new stage, even though the phases are comparable to those in the Wiig model. The phases of the Zack model are acquisition, refinement, storage/retrieval, distribution, and presentation.

The Zack model has been employed to analyze knowledge gaps in marketing and branding contexts (\citet{sihotang2022knowledge}), develop KM strategies through gap and SWOT analysis in Indonesian startups (\citet{budiman2022knowledge}), and align corporate and KM strategies for designing a KM system (\citet{suroso2018designing}).

\textbf{Bukowitz and William Model (2000):} This model improves upon the preceding two by extending the definition of knowledge storage to encompass the infrastructure that supports this learning community (such as communication, hierarchy, and working relationships). Bukowitz and William also stress the importance of gradually expanding your KMS in addition to keeping it up to date. The phases here are: get, use, learn, contribute, and assess.

The Bukowitz and Williams model has been utilized to assess the KM status in medical sciences university libraries in Tehran (\citet{khorami2021status}), identify and rank factors influencing KM implementation in Guilan universities (\citet{taleghani2013identification}), and evaluate KM processes within a research and development organization (\citet{zahedi2021examining}).

\textbf{McElroy Model (2003):} McElroy's focus on knowledge production and integration builds on the Bukowitz and William model's process focus. In an effort to enhance group learning, it provides a mechanism for team members to make "claims" when they do not comprehend or receive information. The phases in the McElroy model are learning, validation, acquisition, integration, and completion.

The McElroy model has been applied in diverse contexts, including addressing communication gaps in the return-to-play process for athletes that suffer injuries (\citet{hazelwood2024optimizing}), enhancing disaster management strategies to mitigate flood damages (\citet{YOUSEFIMOHAMMADI2024100431}), and implementing KM strategies to prevent knowledge loss in government agencies \citet{hlongwane2023use}.

\begin{table}[H]
\begin{scriptsize}
\centering
\setlength{\tabcolsep}{0.5em} 
{\renewcommand{\arraystretch}{2} 
\begin{tabular}{|p{1.8cm}|p{5.3cm}|p{5.3cm}|}
\hline
\textbf{KM Model} & \textbf{Strengths} & \textbf{Challenges} \\
\hline
Wiig Model & 
+ Emphasizes organization and transformation of knowledge, which aligns well with the codification of playbooks.

+ Provides a clear structure for how knowledge is built, stored, pooled, and extended, essential for playbook management. & 
- Lacks emphasis on knowledge sharing and collaboration, critical for incident response scenarios where real-time knowledge exchange is paramount.

- Limited focus on the iterative improvement of knowledge after initial codification. \\
\hline
Zack Model & 
+ Offers a logical and systematic approach, ensuring that playbooks progress through distinct, well-defined phases.

+ Encourages retrieval and presentation strategies, which are crucial for ensuring accessibility during incidents. & 
- Heavily process-driven, which may limit flexibility in scenarios requiring adaptability.

- Limited focus on the iterative improvement of knowledge after initial codification. \\
\hline
Bukowitz and William Model & 
+ Expands knowledge management to include infrastructure and the dynamics of learning communities, promoting better integration with organizational culture.

+ Emphasizes continuous learning and contribution, a key factor in adapting playbooks to evolving threats. & 
- Broad focus on learning communities may dilute the attention needed for tactical, high-priority scenarios like incident response. \\
\hline
McElroy Model & 
+ Introduces mechanisms for knowledge validation and group learning, fostering trust and accuracy in the KMS.

+ Supports claims-based learning, encouraging users to flag inconsistencies or gaps in the playbooks. & 
- Focus on group learning and validation may create bottlenecks, delaying critical updates to playbooks and hindering rapid adaptation to emerging cybersecurity threats. \\
\hline
\end{tabular}
}
\caption{KM Models Comparison in the Context of Security Operations Playbooks}\label{km_models_comparison}
\end{scriptsize}
\end{table}

\section{CACAO Knowledge Management System: Overview and Design}

Although many KM models define structured stages, they primarily serve as guiding frameworks. The effectiveness of knowledge management depends on understanding how each specific use case naturally acquires, retains, and disseminates knowledge (\citet{smartsheet-knowledge-management}). In this paper, we propose a KM model designed to enhance the comprehensive management of CACAO playbooks throughout their lifecycle.

\begin{figure}[h]
\centering
\includegraphics[scale=0.65]{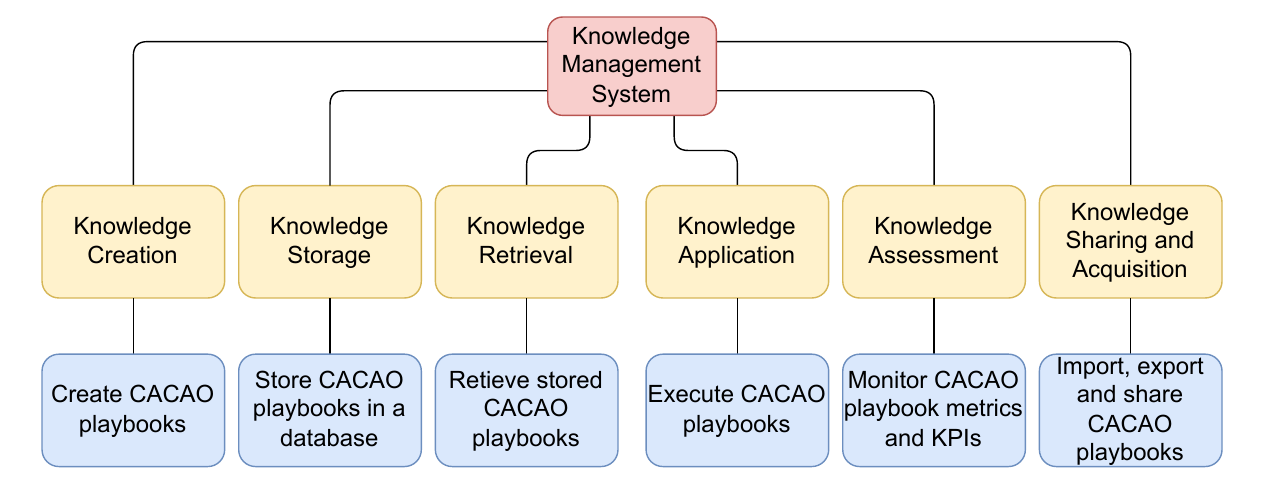}
\caption{Playbook KMS Components \& their mapping to Knowledge Management principles}\label{kb_components}
\end{figure}

The proposed KM model (Figure \ref{kb_components}) for the CACAO playbooks KMS integrates key elements from existing models while addressing their identified challenges in the application domain. It is specifically tailored to align with the unique requirements of cybersecurity operations, ensuring efficient storage, retrieval, and exchange of playbooks:

\begin{enumerate}
  \item Knowledge Creation: Inspired by the Wiig and McElroy models, this stage emphasizes both the codification and validation of knowledge at the point of creation.
  \item Knowledge Storage: Adopts the infrastructure considerations of the Bukowitz and William model, emphasizing secure and scalable storage solutions tailored to playbook data formats.
  \item Knowledge Retrieval: Draws from the Zack model emphasis on systematic retrieval processes, ensuring relevant playbooks are easily accessible during incidents.
  \item Knowledge Application: Expands on the action-oriented “use” phase of the Bukowitz and William Model, focusing on real-time usability.
  \item Knowledge Assessment: Reflects McElroy's “validation” and “learning” phases, ensuring that playbooks are continuously refined based on performance, user feedback and evolving threats.
  \item Knowledge Sharing and Acquisition: Incorporates the collaborative focus of the Bukowitz and William Model to ensure knowledge is shared efficiently across teams, ensuring that playbooks are up-to-date and promoting innovation.

\end{enumerate}

\subsection{System Requirements}
\label{subsec3.1}

The proposed KM model establishes a set of general user requirements for the KMS, ensuring that the application can accommodate diverse user needs while aligning with the strategic objectives of organizations. This approach enables the effective storage, retrieval and use of CACAO playbooks, fostering collaboration and enhancing the cybersecurity operations capability.

\textbf{R1 - Knowledge Creation:} It ensures that users can create new CACAO playbooks.

\textbf{R2 - Knowledge Storage:} It ensures that users can store and manage playbooks in a database, with version control mechanisms to monitor changes.

\textbf{R3 - Knowledge Retrieval:} It ensures that users can retrieve playbooks through advanced search and filtering capabilities, allowing them to find information based on various criteria and providing fast and efficient access to relevant knowledge.

\textbf{R4 - Knowledge Application:} It ensures that users can execute the stored playbooks via the support of dedicated software for orchestration and automation.

\textbf{R5 - Knowledge Assessment:} It ensures that users will be able to monitor playbook performance through metrics and key performance indicators (KPIs), providing information on their effectiveness and areas for improvement.

\textbf{R6 - Knowledge Sharing and Acquisition:} It ensures that the system supports knowledge sharing by allowing the user to import, export, and share CACAO playbooks. In addition, the authenticity of a playbook's signature can be verified.

\subsection{System Architecture}

\begin{figure}[h]
\centering
\includegraphics[scale=0.65]{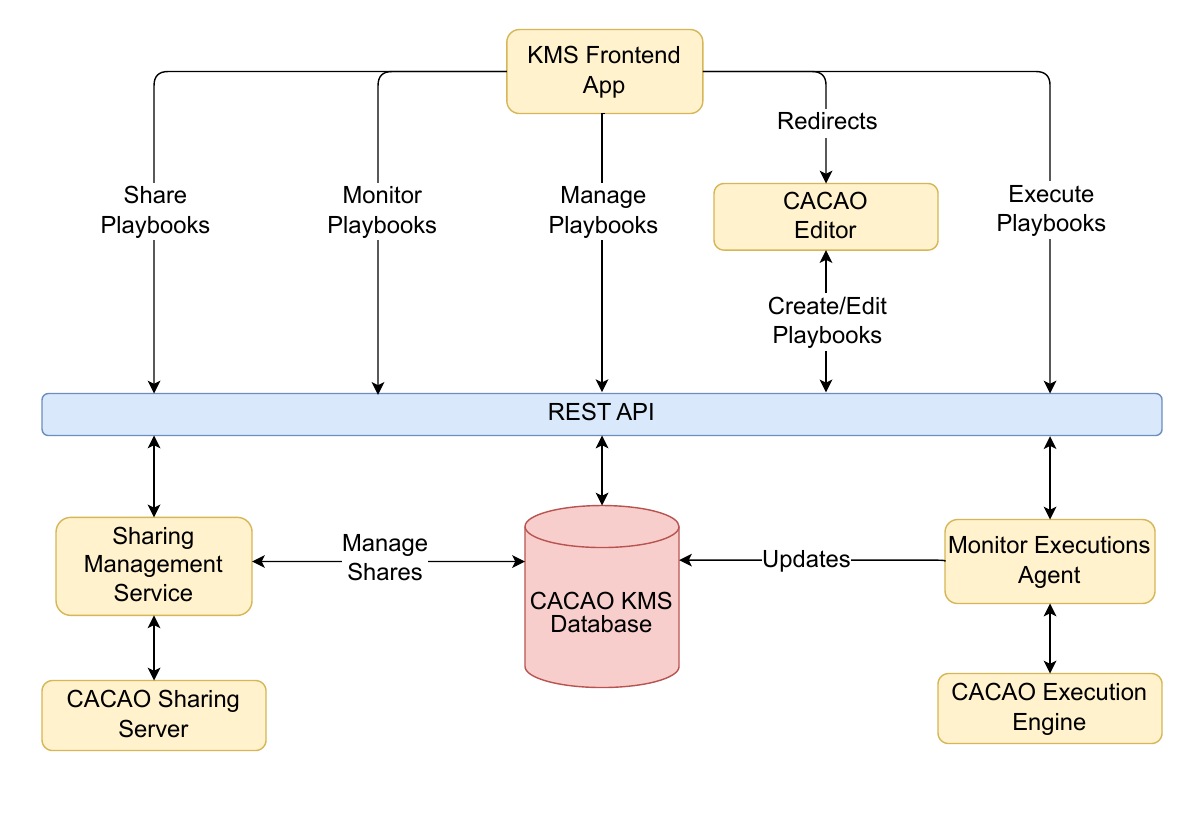}
\caption{General System Architecture}\label{general_sys_arch}
\end{figure}

The proposed system architecture (Figure \ref{general_sys_arch}) for the CACAO playbooks KMS comprises multiple components designed to facilitate the creation, management, execution, sharing, and monitoring of playbooks. At its core, the architecture is built around a REST API, which enables seamless communication between the frontend, various services, and underlying databases. The key components of the system are detailed below, outlining their roles in ensuring efficient and scalable playbook management.

\textbf{Frontend:} This main user interface allows end users to interact with CACAO playbooks. The frontend application offers the following functions:

\begin{enumerate}
  \item Managing playbooks
  \item Creating and editing playbooks
  \item Executing playbooks
  \item Sharing playbooks
  \item Monitoring playbook metrics
\end{enumerate}

\textbf{CACAO Editor:} This component enables users to create and modify CACAO playbooks through an intuitive, user-friendly interface. When users perform actions such as creating or editing a playbook, the system communicates with the REST API to store or update the corresponding data in the database. Additionally, the Editor integrates built-in functionalities that support other services, including playbook execution and sharing, enhancing the overall usability and efficiency of the system.

\textbf{REST API:} The REST API serves as the central communication hub for the entire system, managing requests from the frontend application, the CACAO Editor, and other supporting services. By abstracting the complexity of internal operations, it ensures seamless, standardized, and secure interactions between system components, facilitating efficient data exchange and integration.

\textbf{Database:} The database serves as the central repository for all information related to CACAO playbooks, fulfilling the general requirement R2 - Knowledge Storage. It plays a crucial role in preserving the integrity of playbooks while also recording actions and various operational statistics. Additionally, it maintains older versions of playbooks, allowing users to review, track changes, and restore previous iterations as needed.

\textbf{CACAO Execution Engine:} The system also incorporates the capability to execute CACAO playbooks, enabling users not only to store and manage them but also to run them directly. This integration ensures that playbooks can be tested efficiently, allowing users to assess their functionality and performance under real-world conditions, thereby enhancing their effectiveness in cybersecurity operations.

\textbf{Monitor Executions Agent:} This agent is responsible for tracking the execution progress of CACAO playbooks. It receives updates from the CACAO Execution Engine and transmits them to the KMS database. These updates are then displayed to users through the Execute Page of the frontend application, ensuring transparency and enabling real-time monitoring of playbook execution.

\textbf{CACAO Sharing Server:} The Playbook Sharing Server facilitates the secure distribution and sharing of CACAO playbooks with external systems and users. It operates alongside the Sharing Management Service to ensure that playbooks are shared efficiently.

\textbf{Sharing Management Service:} This service oversees the controlled sharing of playbooks across various platforms and users. It interacts with the CACAO Sharing Server to ensure secure distribution while maintaining a controlled sharing environment. Additionally, it synchronizes sharing activities with the database, ensuring that users can always access the most up-to-date shared content.

\section{Proof of Concept}
\label{sec4}

\subsection{Implementation - Core Components}

\begin{figure}[h]
\centering
\includegraphics[scale=0.65]{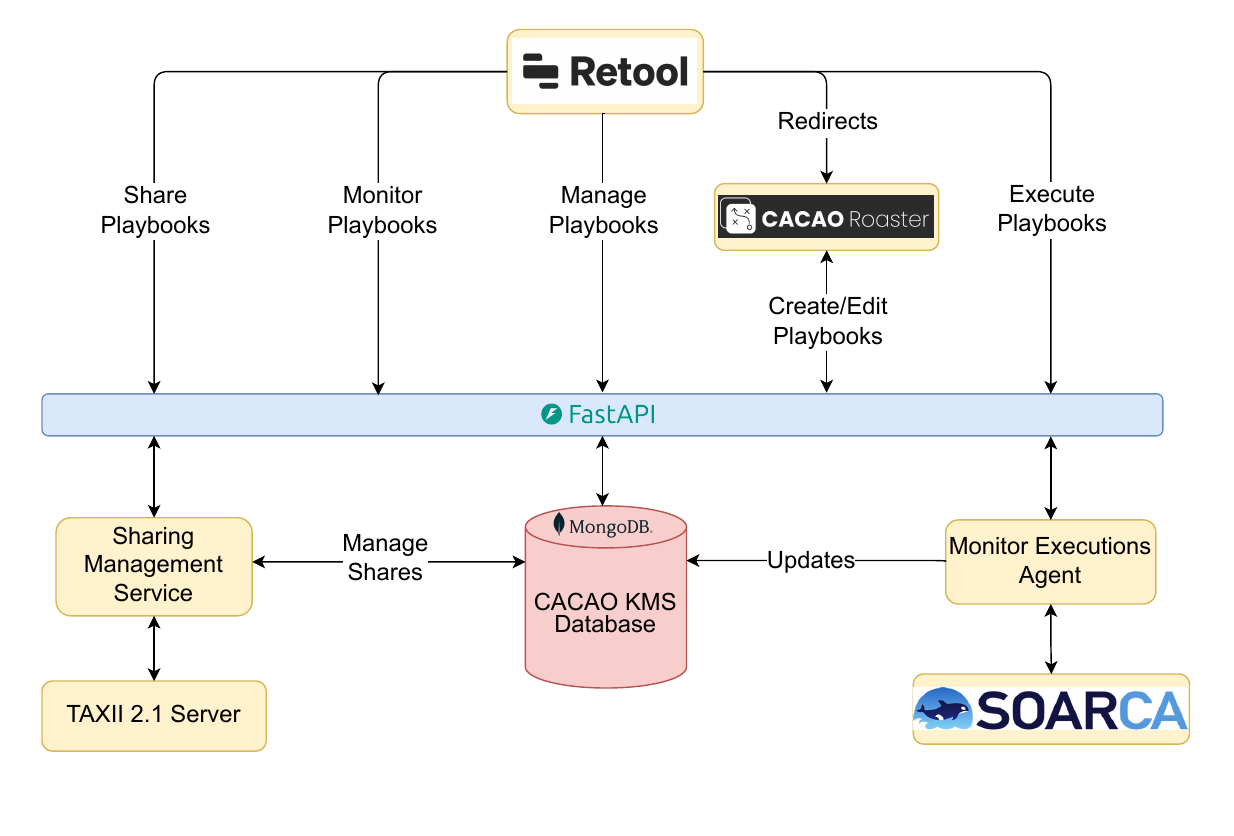}
\caption{System Architecture}\label{sys_arch}
\end{figure}

In this section, we outline the platform's development process and the implementation of its core functionalities. Figure \ref{sys_arch} provides a detailed system architecture diagram, showcasing the technologies employed for each component and their interactions within the system.

\textbf{Frontend:} Retool was selected for the development of the KMS user interface due to its robust low-code capabilities, which streamline the creation of internal tools and systems. This platform enables developers to efficiently build applications with minimal coding effort while offering a powerful environment for designing, developing, and managing complex data-driven applications. Additionally, Retool simplifies the integration process and seamlessly connects with various databases and APIs, enhancing the overall efficiency and scalability of the system (\citet{retool-website}).

Retool's ability to rapidly and intuitively create user interfaces aligns with the need for swift prototyping and application development, particularly in academic and research settings. Additionally, its advanced data management capabilities ensure that the KMS can effectively handle the complex data interactions necessary for managing CACAO playbooks throughout their lifecycle.

\textbf{CACAO Editor:} CACAO Roaster was chosen as the playbook editor for the system, as it is a web-based application developed under the Open Cybersecurity Alliance (OCA) initiatives. This tool provides an integrated environment for creating, analyzing, validating, manipulating, and visualizing CACAO playbooks. Its primary objective is to simplify the process of playbook creation and management, offering a user-friendly interface that does not require extensive programming expertise (\citet{cacao-roaster-github}).

Its key features include:

\begin{enumerate}
  \item Creating and editing playbooks: Users can create new playbooks or edit existing ones using a graphical user interface, greatly simplifying the process.
  \item Validation: The application includes the ability to validate playbooks and their signatures to ensure their authenticity and that they comply with the formal technical specification.
  \item Visualization: Visual representation of playbooks makes it easier to understand their functionality, assisting in their creation and modification, as well as when exchanged between cybersecurity teams.
  \item Integration: CACAO Roaster integrates with other projects such as STIX 2.1 and OpenC2 (Open Command and Control), promoting interoperability and integrated threat response strategies.
\end{enumerate}

To integrate CACAO Roaster into the KMS, the system was designed to enable the Editor to access stored playbooks within the database, allowing users to retrieve, modify, and save them seamlessly. Additionally, enhancements were introduced to extend the editor’s functionality, enabling users to execute and share playbooks, as well as restore previous versions when necessary.

\textbf{REST API:} FastAPI was selected for the implementation of the REST API due to its advanced asynchronous capabilities and superior performance. Unlike traditional frameworks such as Flask and Django, FastAPI leverages Python's asynchronous features, resulting in improved efficiency and scalability, particularly when managing concurrent requests (\citet{capitalnumbers-django-flask-fastapi}). Furthermore, FastAPI automatically generates API documentation using OpenAPI, streamlining the development process and improving productivity (\citet{fastapi}).

\textbf{Database:} MongoDB (\citet{mongodb-introduction}) was selected as the optimal choice for storing CACAO playbooks in JSON format due to its design as a document-oriented NoSQL database, which natively supports JSON data structures. This compatibility ensures seamless integration and efficient search of playbooks without the need for complex transformations. In addition, MongoDB's powerful query capabilities, horizontal scalability and high availability features provide the performance and reliability needed to manage and access large volumes of playbook data, making it an excellent choice for applications that require dynamic, scalable and efficient storage solutions (\citet{mongodb-definitive-guide}).

\begin{table}[h]
\begin{scriptsize}
\centering
{\renewcommand{\arraystretch}{2} 
\begin{tabular}{|>{\centering\arraybackslash}p{1.8cm}|>{\centering\arraybackslash}p{1.8cm}|>{\centering\arraybackslash}p{1.8cm}|>{\centering\arraybackslash}p{1.8cm}|>{\centering\arraybackslash}p{1.8cm}|>{\centering\arraybackslash}p{1.8cm}|}
\hline
\textbf{Approach} & \textbf{Current Query Speed} & \textbf{History Query Speed} & \textbf{Maintaining Versions Complexity} & \textbf{Consistency - Inserting New Versions} & \textbf{Unlimited Growth of Versions} \\
\hline
New doc per version & \textcolor{red}{Slow} & Fast & Low & Strong & Yes \\
\hline
New doc per version inc "current" & Fast & Fast & Medium & \textcolor{red}{Weak} & Yes \\
\hline
Versions embedded in single doc & Fast & Medium & Low & Strong & \textcolor{red}{No} \\
\hline
Document Versioning Pattern & Fast & Medium & Low & Medium & Yes \\
\hline
Deltas only in new doc & \textcolor{red}{Slow} & \textcolor{red}{Slow} & Medium & Strong & Yes \\
\hline
\end{tabular}
\caption{Comparison of Version Control Patterns (Source: \citet{askasya-revisit-versions})}\label{version_control_patterns}
}
\end{scriptsize}
\end{table}

For Version Control, a method known as Document Versioning Pattern was chosen from the patterns in Table \ref{version_control_patterns}, in which a separate collection of documents is maintained for the latest versions of the playbooks and another for the history ones. The above method is considered the best choice for the following reasons:

\begin{enumerate}
  \item Each document does not have too many versions.
  \item Documents are not updated too often, as not too many consecutive playbook updates are expected.
  \item Most of the queries that are executed are performed on the most recent version of each document, i.e., the newest version of each playbook.
  \item Current data and historical data are usually searched separately.
\end{enumerate}

Therefore, the Document Versioning Pattern allows the system to store the latest playbooks and their history in the same database and also avoids the use of multiple systems for managing the historical data (e.g., git), which increases the complexity of the application (\citet{mongodb-document-versioning}).

\textbf{CACAO Execution Engine:} The SOARCA execution engine was selected for running CACAO playbooks. SOARCA offers an API for its Executor Module, which has been integrated into the KMS, enabling users to execute CACAO playbooks and carry out the predefined commands efficiently.

\textbf{Monitor Executions Agent:} To manage playbook executions and their history, we created a dedicated collection in the database where execution progress from the SOARCA Executor is stored and continuously updated. A monitoring agent was implemented to track and update execution progress. This agent ensures real-time monitoring of execution status and logs key statistics such as execution time and completion state in the database.

The Monitoring Agent operates as an asynchronous background task within the API, leveraging a pull-based mechanism. It is triggered whenever a playbook execution begins and periodically queries the execution status using the SOARCA Reporter Module (\citet{soarca-reporting}) via the SOARCA API. This enables the system to remain aware of whether an execution is ongoing, successfully completed, or has failed, ensuring comprehensive tracking and visibility into playbook execution history.

\textbf{CACAO Sharing Server:} To enable seamless and secure sharing of CACAO playbooks, the Trusted Automated eXchange of Indicator Information (TAXII 2.1) protocol (\citet{taxii-v2.1}) was selected for integration within the KMS. By combining TAXII 2.1 with the Structured Threat Information eXpression (STIX 2.1) standard (\citet{stix-v2.1}), the system ensures interoperability and enhances collective defense efforts by facilitating structured, automated threat intelligence sharing.

STIX 2.1—developed by OASIS—provides a standardized language for representing and exchanging CTI in a machine-readable format. It enables organizations to describe various cybersecurity aspects, including indicators, attack patterns, threat actors, and mitigation techniques. Given its purpose, high scalability, and flexibility, STIX 2.1 is well-suited for integrating with CACAO playbooks.

A key feature of STIX 2.1 is the CoA object, which defines specific response actions for identified threats. This aligns directly with CACAO playbooks, as COAs can encapsulate the structured response steps required to address cyber incidents. By leveraging these standards, the KMS ensures that playbooks can be effectively shared, understood, and applied across different organizations and security ecosystems (\citet{stix-v2.1}).

\begin{figure}[h]
\centering
\includegraphics[scale=0.85]{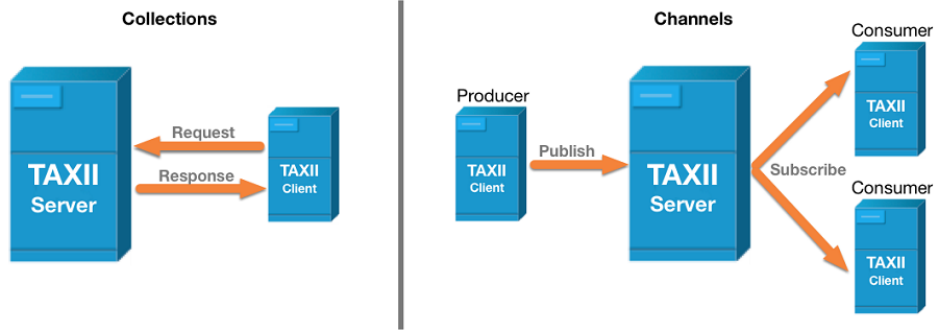}
\caption{TAXII 2.1 Sharing Models}\label{sharing_models}
\end{figure}

TAXII 2.1 is a protocol designed to enable the secure exchange of CTI via HTTPS. TAXII provides a set of services and message exchanges for CTI sharing, typically represented using the STIX format. A TAXII 2.1 server acts as a repository where organizations can publish, store, and retrieve STIX objects, which can also incorporate CACAO playbooks. These servers are designed to be scalable and secure, supporting controlled distribution of CTI between organizations. 

One of the key features of TAXII 2.1 is its ability to facilitate the sharing of STIX objects through Collections. A Collection is an interface to a logical repository of CTI objects provided by a TAXII Server that allows a producer to host a set of CTI data that consumers can request.

In TAXII there are two sharing models, the Collections model and the Channels model (Figure \ref{sharing_models}). In the Collections model, each playbook is represented as a STIX 2.1 CoA object included in a TAXII Collection and can be exchanged in a request-response model. In the Channels model, organizations can subscribe to these Collections to receive updates or new playbooks as they are published. This ensures that all participating organizations have access to the latest and most up-to-date playbooks (\citet{taxii-v2.1}).

In the 2.1 version of the specification, Channel services are not specified and will be defined in a later version of TAXII. For this reason, we connected the KMS to a TAXII 2.1 server with the Collections model, which is supported by a RESTful API, making it easy to integrate with other systems and automate knowledge sharing. To provide an easier and more user-friendly sharing process, we created the Sharing Management Service to keep track of the playbook sharing actions.

The extension proposed in the research (\citet{mavroeidis2022cybersecurity}) was used to convert CACAO playbooks into STIX Course of Action type objects. This research implements the STIX 2.1 representation of the CACAO playbook, incorporating the metadata standard proposed by research (\citet{mavroeidis2021integration}), thus ensuring that the necessary information for describing, integrating and sharing the playbooks is gathered in the STIX object, allowing for seamless integration and use in existing cybersecurity systems.

\textbf{Sharing Management Service:} To manage playbook sharing, we implemented a new Sharings Collection in the Database, which tracks versions of playbooks that have been shared to or from the TAXII Server.

This Sharings Collection serves multiple purposes:

\begin{enumerate}
    \item It maintains a history of all playbook shares, ensuring traceability.
    \item It prevents duplication, ensuring that each version of a playbook is shared only once.
    \item It enables users to track their shared playbooks, reducing confusion and improving version control.
\end{enumerate}

By leveraging this collection, the system provides structured and controlled dissemination of CACAO playbooks, ensuring efficient knowledge sharing without compromising organization or security.

\subsection{User Actions \& Action Flows}

\begin{figure}[t]
\centering
\includegraphics[scale=0.57]{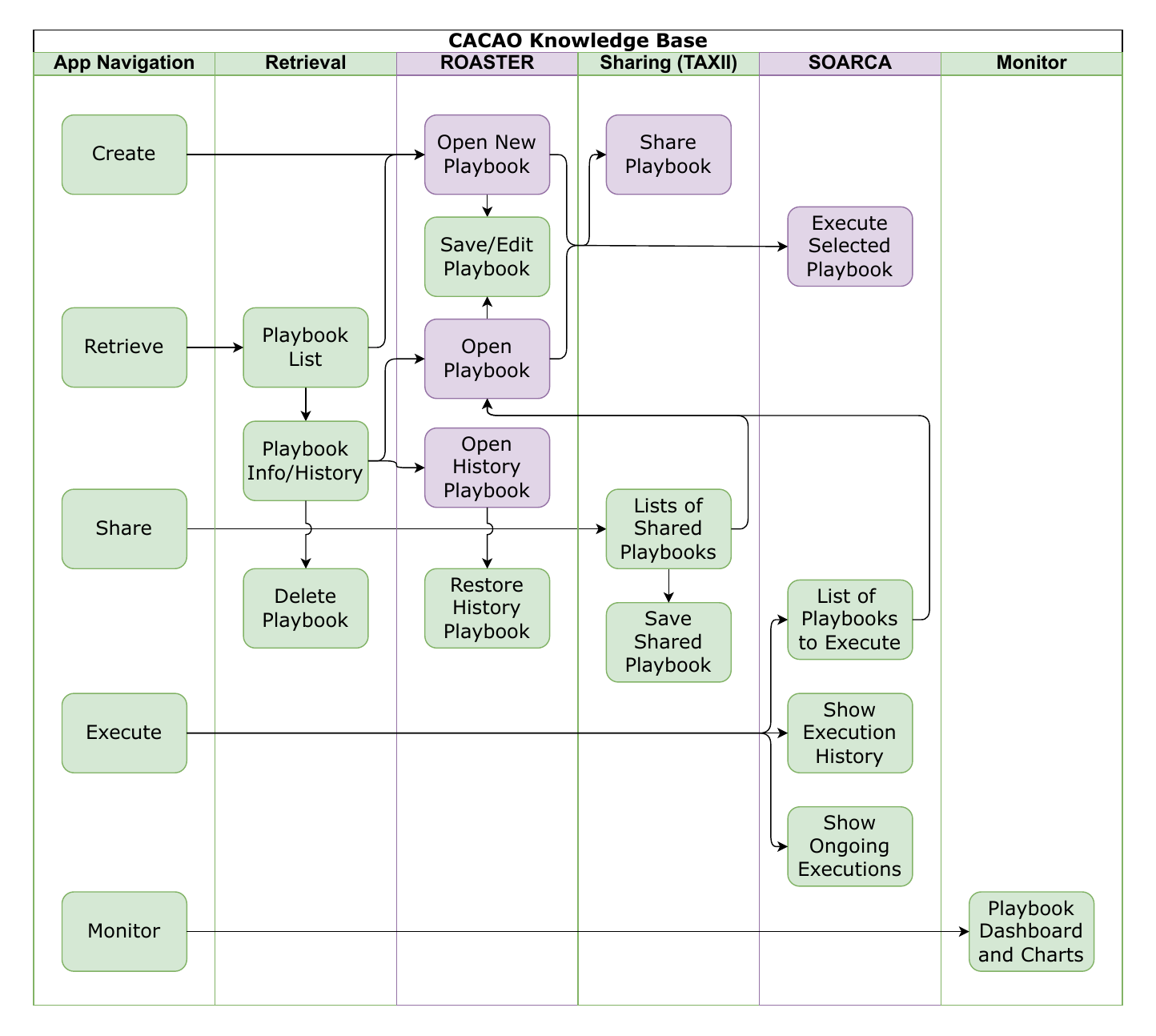}
\caption{Flow Diagram (external tools integrated appearing in purple)}\label{flow_diagram}
\end{figure}

Based on the requirements set in Section \ref{subsec3.1}, we identify the following user actions that define the overall application workflow and present them in Figure \ref{flow_diagram}.

\textbf{Create:} The "Create" page allows users to create a new CACAO playbook. By selecting it, the user is able to open the CACAO Roaster Editor and choose how they want to create the playbook. Once a playbook is created, users can save or edit it as they wish, as well as execute it, share it on the TAXII Server and perform any operation offered by the Editor. Meets the general requirement R1 - Knowledge Creation.

\textbf{Retrieve:} The "Retrieve" page allows users to retrieve existing playbooks, view their main properties and manage them. Access is given to a list, which contains all saved CACAO playbooks, from which users can view detailed information about each one and monitor its history, displaying previous versions and changes. The user is able to perform management operations on each playbook, such as deleting it, opening it in the CACAO Roaster Editor, as well as opening a previous version, which can be restored via the Editor. Meets the general requirement R3 - Knowledge Retrieval.

\textbf{Share:} The "Share" page allows users to share their own saved playbooks or save new ones shared by external parties. Access is given to a list of existing playbooks that can be shared on the TAXII Server, as well as a list of CACAO playbooks that have already been shared on the TAXII Server and can be saved by the user. Meets the general requirement R6 - Knowledge Sharing and Acquisition.

\textbf{Execute:} The "Execute" page allows users to execute an existing playbook, as well as view available information about the executions. Access is given to a list of existing playbooks that can be sent to the SOARCA system and be executed, a list containing the executions that are currently in progress, and a list of the history of the executions and various information about them. Meets the general requirement R4 - Knowledge Application.

\textbf{Monitor:} The "Monitor" page allows users to monitor and check the system’s operation. This page offers a dashboard that gathers data about available playbooks and their executions, displaying database information and other historical data. It also provides information on the overall effectiveness of the playbooks through various charts and KPIs, from which users can gain important insights into their functionality. Meets the general requirement R5 - Knowledge Assessment

\subsection{Proof of Concept Showcase}

\subsubsection{Demonstration}

In this section, we present a detailed demonstration of the platform through a series of screenshots, showcasing its key features and functionalities. In addition, we provide a link to a demo video that offers a comprehensive walkthrough of the system in action. The video can be accessed at \href{https://www.youtube.com/watch?v=6fGhg02aMlg}{this link}.

\begin{figure}[H]
\centering
\includegraphics[scale=0.45]{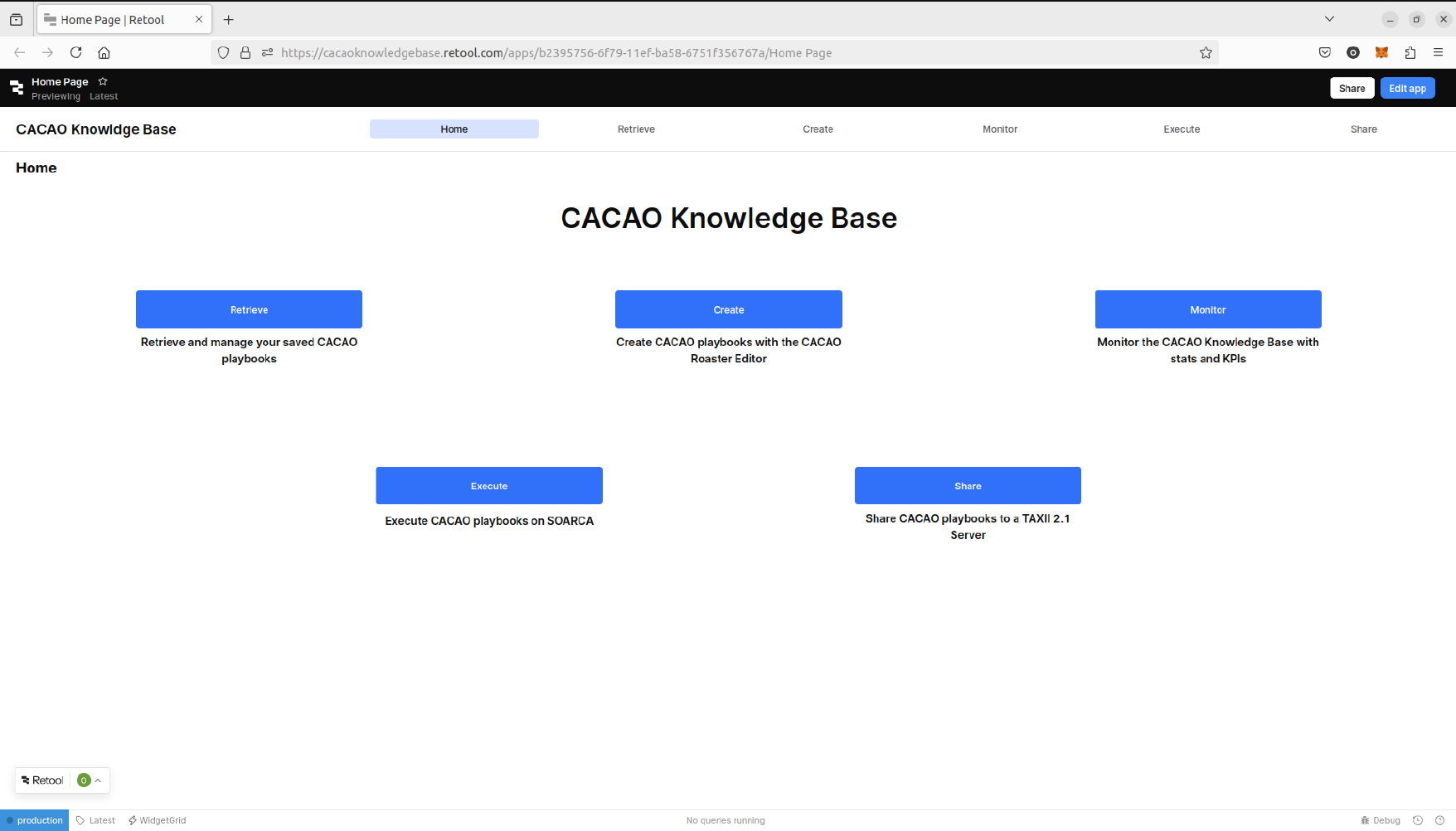}
\caption{The Playbook KMS Landing Page}\label{landing_page}
\end{figure}

Figure \ref{landing_page} illustrates the Landing Page of the application, which serves as the main navigation hub. It contains five primary buttons: Retrieve, Create, Monitor, Execute, and Share, each representing a core functionality of the system.

\begin{figure}[H]
\centering
\includegraphics[scale=0.45]{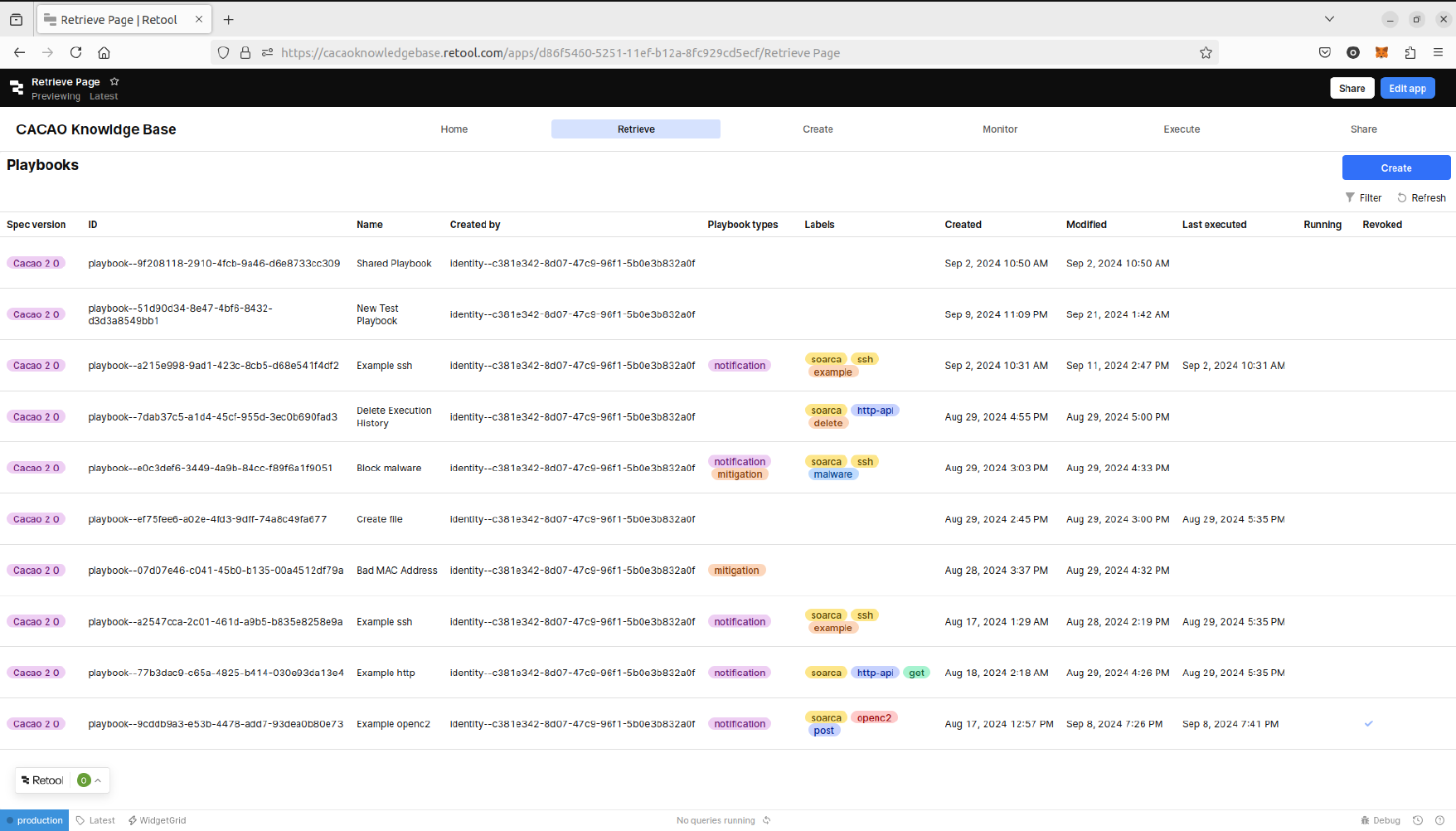}
\caption{Retrieve Page}\label{retrieve_page}
\end{figure}

Figure \ref{retrieve_page} presents the Retrieve Page, which allows users to access a list of all saved playbooks. Here, users can view basic information about the playbooks, such as the ID, name, creator, creation and modification dates, etc.

\begin{figure}[H]
\centering
\includegraphics[scale=0.45]{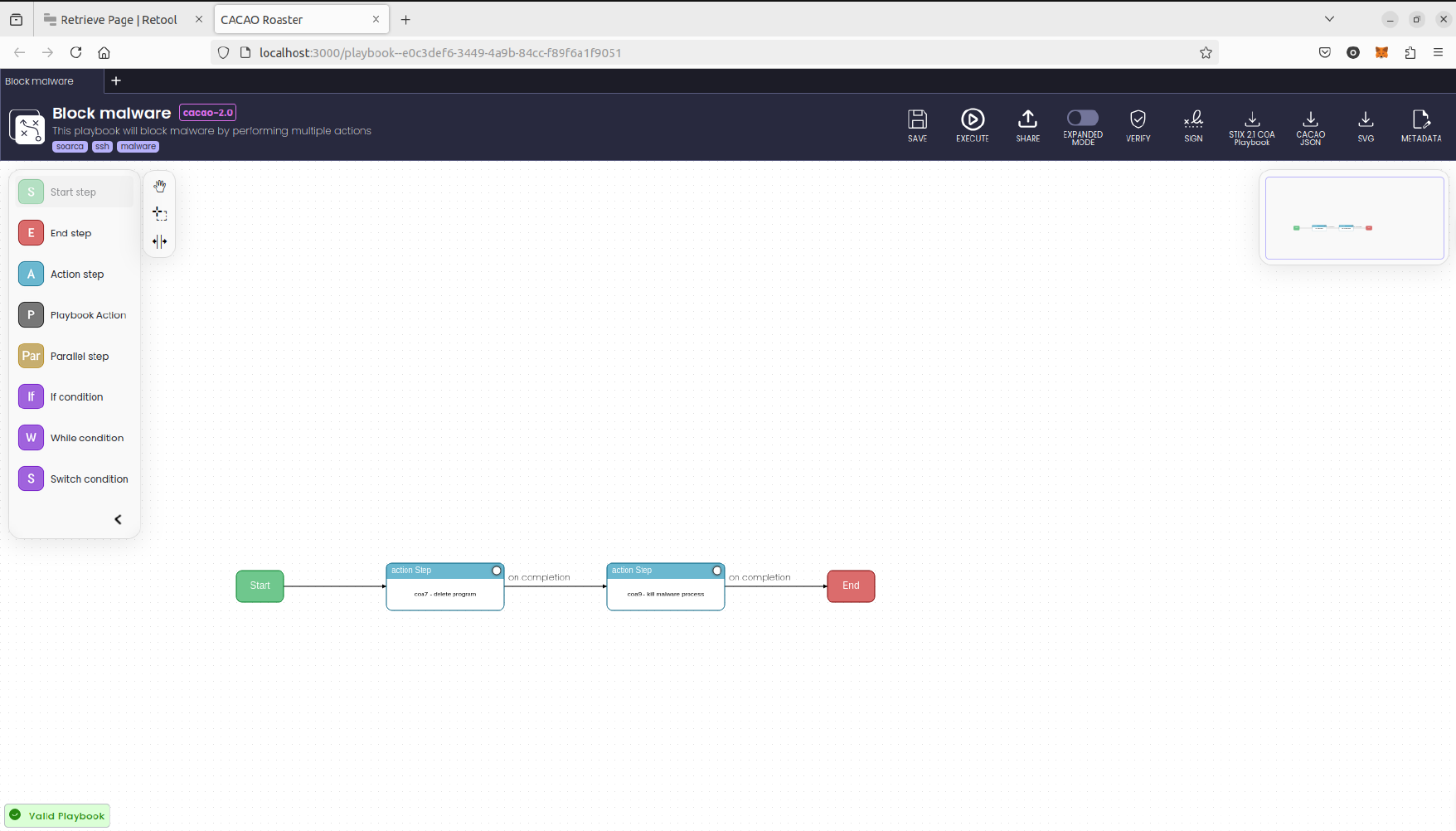}
\caption{Create Page - CACAO Roaster}\label{create_page}
\end{figure}

Figure \ref{create_page} demonstrates the Create Page, which allows users to create playbooks through the CACAO Roaster Editor.

\begin{figure}[H]
\centering
\includegraphics[scale=0.45]{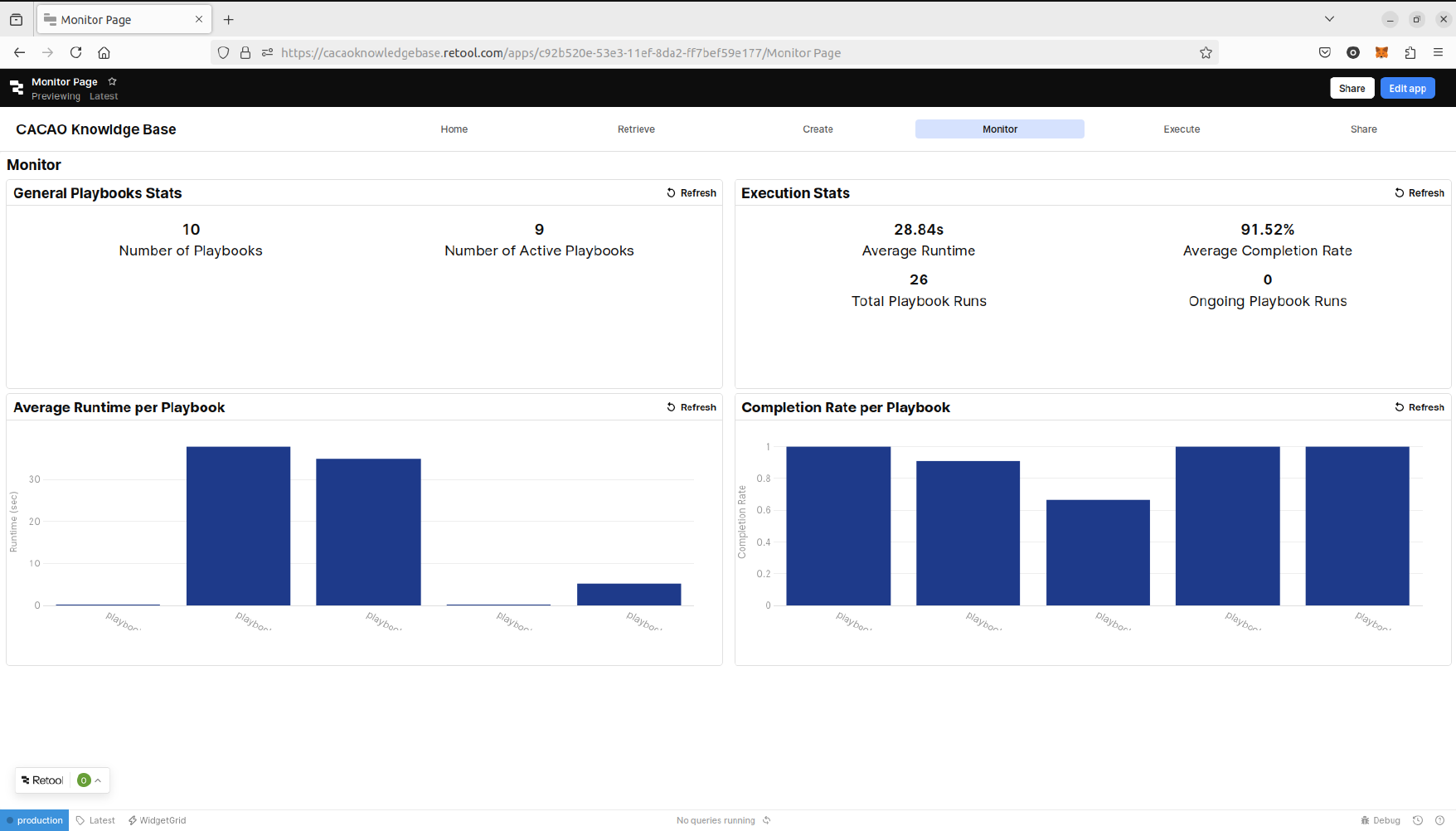}
\caption{Monitor Page}\label{monitor_page}
\end{figure}

Figure \ref{monitor_page} shows the Monitor Page, which allows users to monitor the statistics and performance of the saved playbooks. This page contains tables and charts with KPIs, which provide information about the platform and its effectiveness.

\begin{figure}[H]
\centering
\includegraphics[scale=0.45]{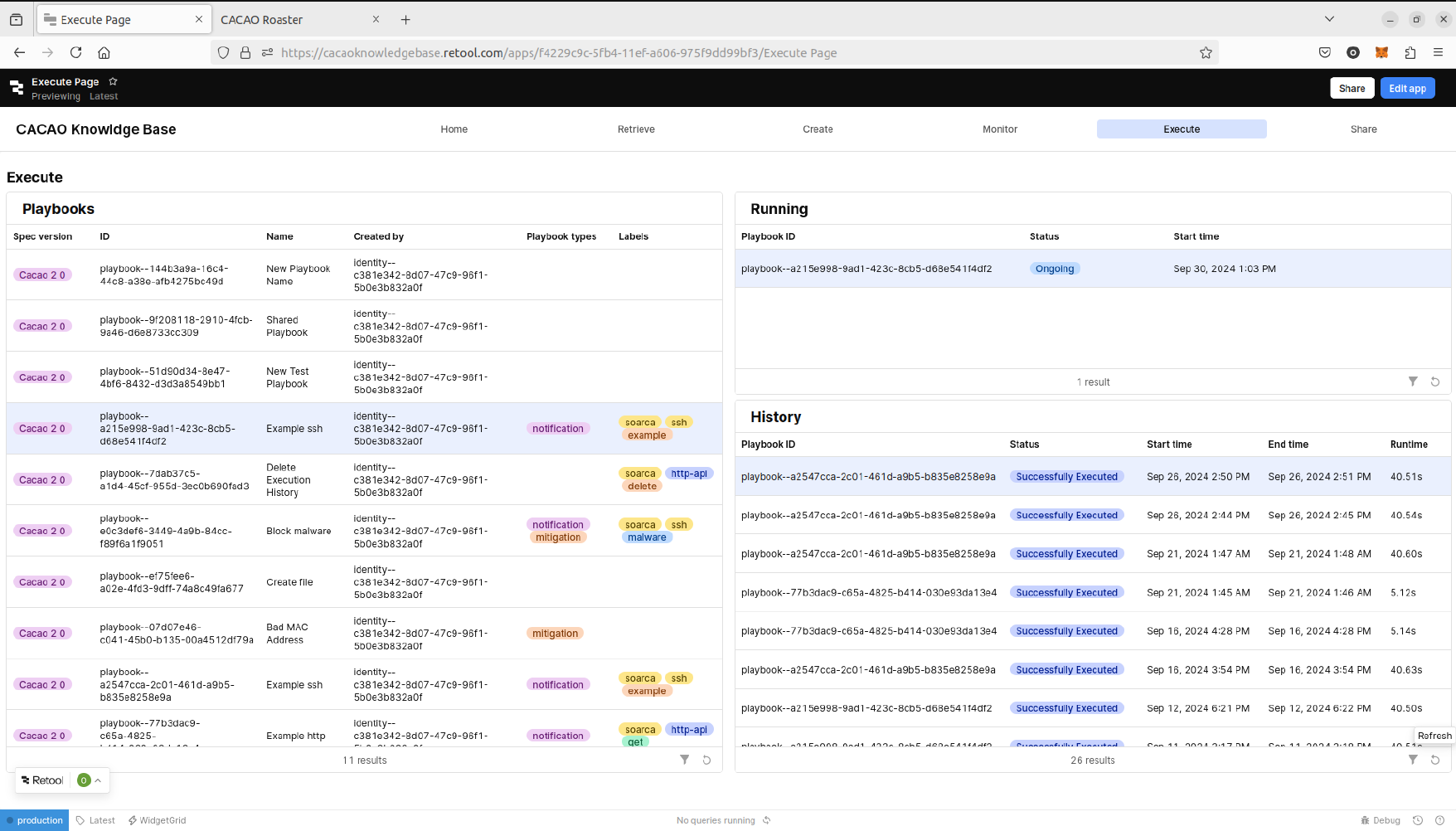}
\caption{Execute Page}\label{execute_page}
\end{figure}

Figure \ref{execute_page} demonstrates the Execute Page, which allows users to trigger the execution of a playbook and monitor its progress.

\begin{figure}[H]
\centering
\includegraphics[scale=0.45]{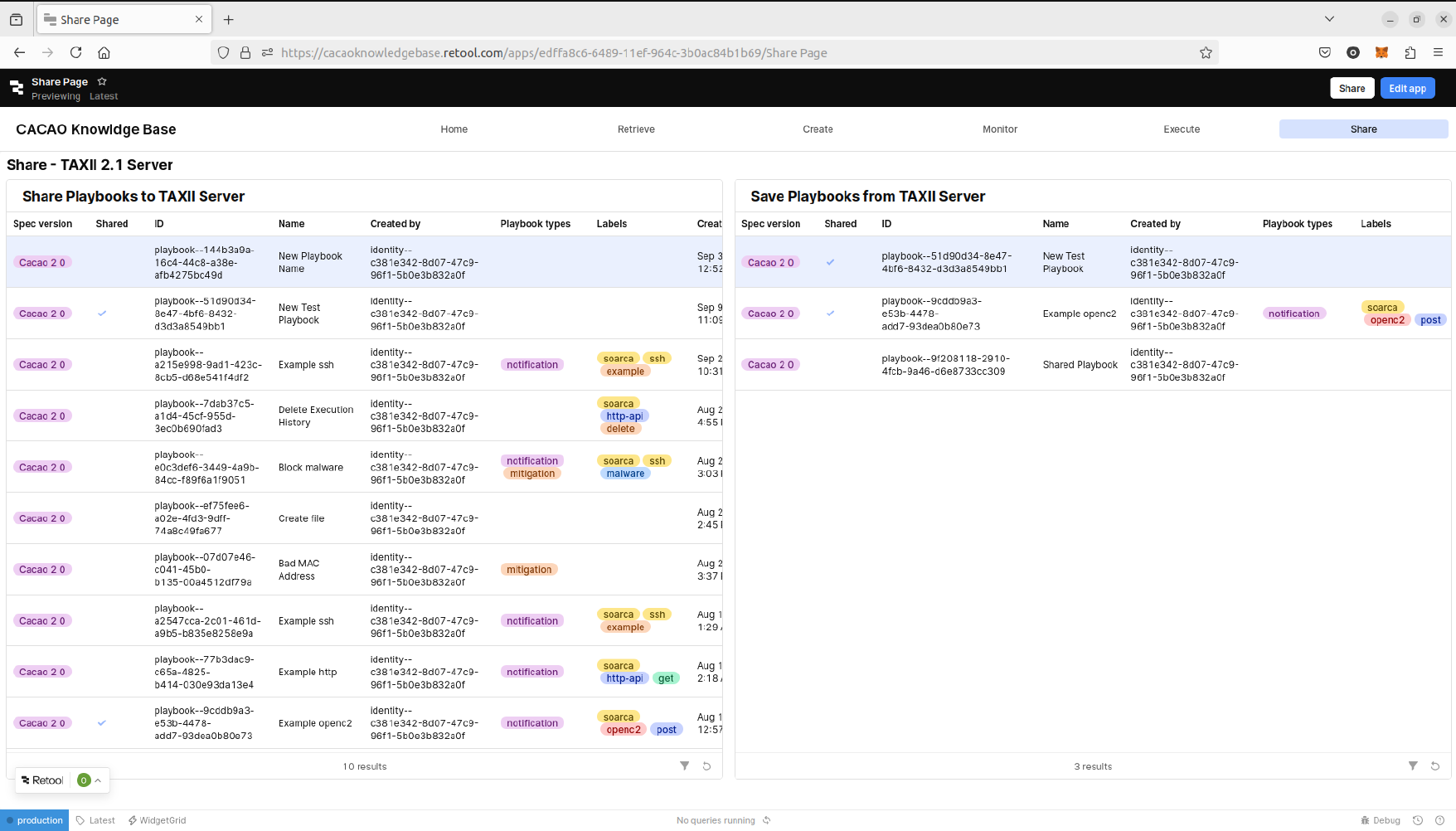}
\caption{Share Page}\label{share_page}
\end{figure}

Figure \ref{share_page} illustrates the Share Page, which allows for secure sharing of playbooks between different organizations or individuals.

\subsection{Evaluation}

The KMS was designed to provide an efficient and scalable infrastructure for storing and managing the CACAO playbook. This section evaluates the solution based on its compliance with the CACAO specification, API performance, memory and storage requirements, scalability, and overall system performance. \footnote{The evaluation was performed on an Ubuntu 22.04.4 LTS system, with an AMD Ryzen 5 3600 6-Core Processor @3.6GHz and 16 GB RAM.} \\

\textbf{API performance}

The performance of the API was assessed using Apitally, an API performance monitoring tool that provides detailed insights into key metrics such as response times, throughput, and latency (\citet{apitally}). The evaluation involved two traffic load scenarios: (1) a baseline test with a single user operating under normal conditions and (2) a stress test simulating approximately 60 users under heavier load. Each test ran for a duration of 10 minutes, allowing for a comparative analysis of system behavior under varying levels of demand.

\begin{figure}[h]
\centering
\includegraphics[scale=0.41]{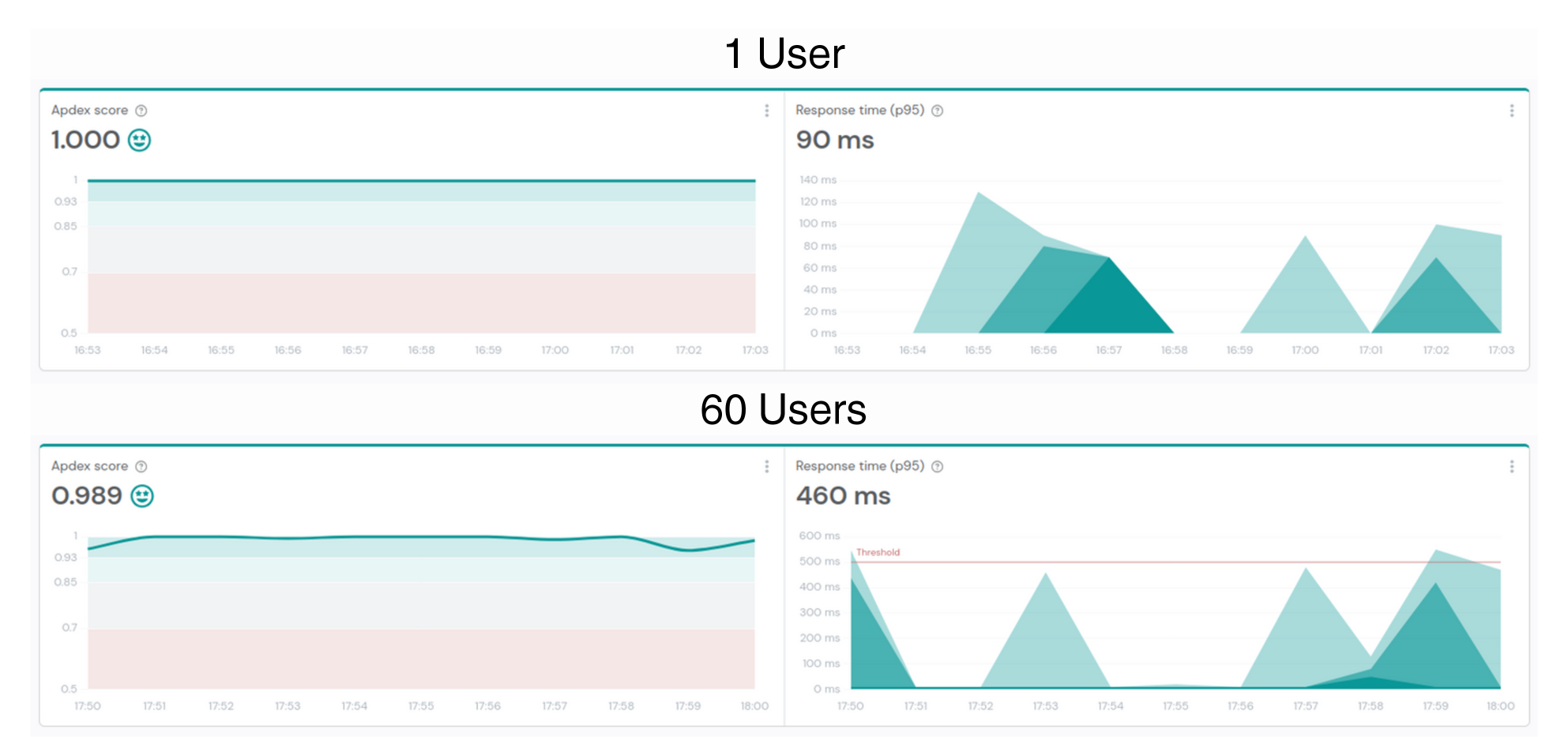}
\caption{API Response Times}\label{response_times}
\end{figure}

The response time of the API was evaluated, measuring the average response times for key endpoints such as playbook creation, retrieval, and deletion. Under normal conditions, the 95th percentile response time across all endpoints was recorded at 90 milliseconds. Under heavier load, with approximately 60 concurrent users, this value increased to 460 milliseconds, as illustrated in Figure \ref{response_times}. These results indicate that the system maintains satisfactory performance, ensuring responsiveness suitable for real-time applications.

The system's throughput was also evaluated. Under normal usage, the system processed an average of 9 requests per minute. When subjected to higher loads with an increasing number of concurrent users, it successfully handled up to 548.73 requests per minute without significant degradation in performance. This demonstrates the system's ability to scale efficiently while maintaining stable throughput.

\begin{figure}[h]
\centering
\includegraphics[scale=0.41]{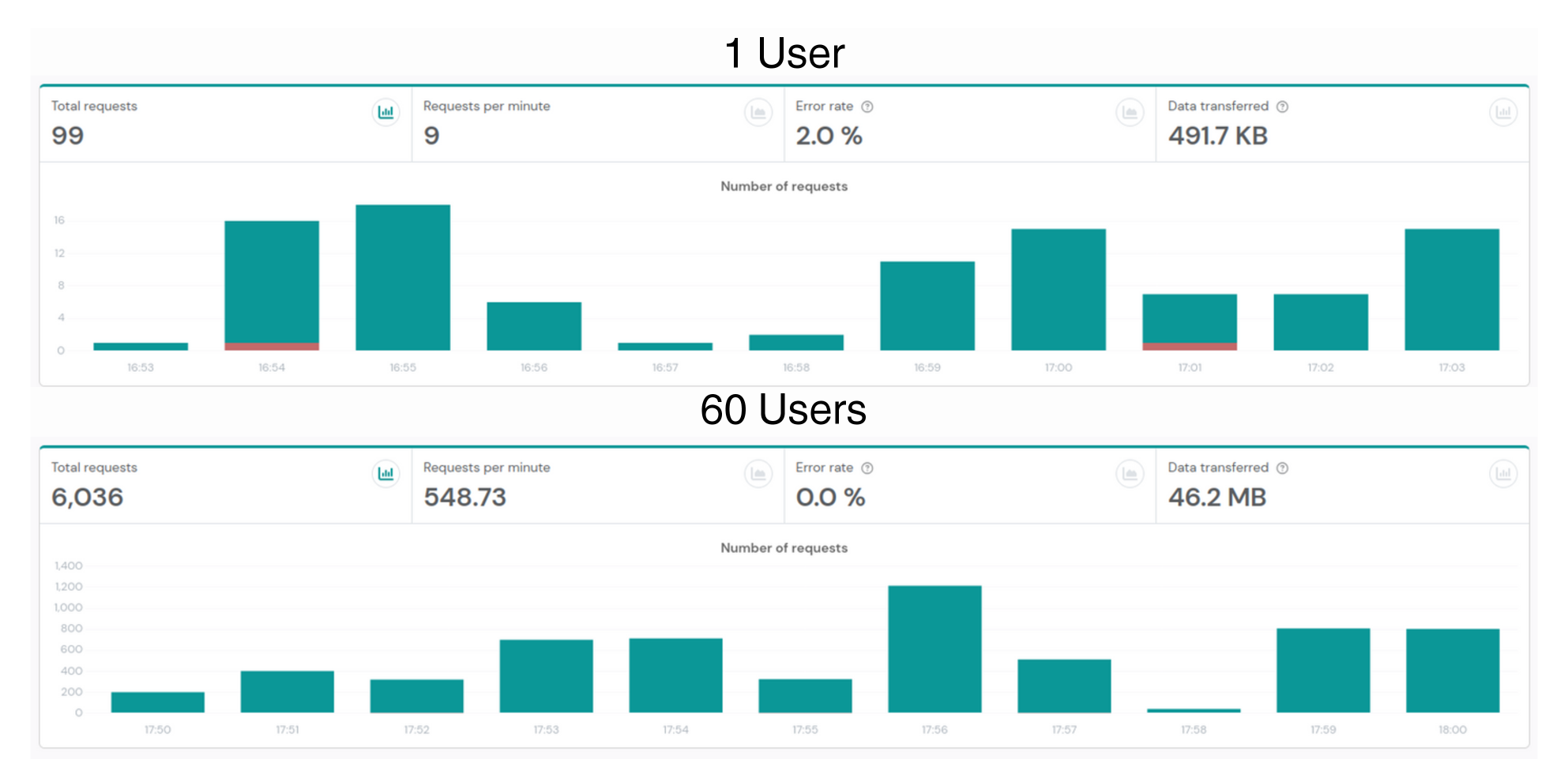}
\caption{API Requests Stats}\label{requests_stats}
\end{figure}

The API latency remained stable under typical conditions but showed a minimal increase during peak traffic periods. The error rate remained consistently low at below 2.0\%, demonstrating the API’s high reliability even under heavy usage. The observed errors were primarily related to attempts to delete non-existent sharing data when removing playbooks. Additionally, data transfer analysis revealed that 46.2 MB of data was transferred during the high load period, compared to 491.7 KB during normal usage, highlighting the API's capacity to handle substantial data volumes efficiently, as shown in Figure \ref{requests_stats}.

\textbf{Memory requirements}

Running the backend of the application, including the TAXII 2.1 Server, requires approximately 2.75 GB of RAM under normal operation and does not exceed 3 GB even under high usage conditions.

For the demonstration, relatively simple playbooks were used, with an average document size of 3.2 KB. However, in real-world scenarios where more complex playbooks are stored, the document size is expected to increase to approximately 50-100 KB per playbook. Considering a medium-sized organization that maintains around 100 active playbooks, each with 10 versions, the estimated storage requirement would range from 50 MB to 100 MB.

\textbf{Scalability}

Ensuring the scalability of the CACAO playbook KMS is crucial as the volume of playbooks and active users grows. Scalability in this context refers to the system's ability to handle increasing demands on storage, retrieval, and execution without significant slowdowns. A rise in the number of playbooks or concurrent users could lead to performance bottlenecks due to higher processing and memory requirements. However, by leveraging a well-structured database architecture—such as distributed databases (sharding) and efficient indexing, which MongoDB supports—the system can mitigate latency issues in searching, sharing, and executing playbooks. Additionally, implementing caching mechanisms at the API level can significantly improve performance, particularly for frequently accessed data. Continuous performance monitoring as playbook volume grows will be essential to detect and address scalability challenges proactively.

\textbf{Compliance with CACAO standard}

The developed system is fully compliant with the CACAO 2.0 standard, meeting all mandatory requirements outlined in Section 11 (Compliance) of the specification. CACAO defines a set of mandatory and optional elements for the implementation and management of playbooks, which all software systems using it are required to follow. By fully aligning with CACAO 2.0, the system guarantees interoperability, security, and structured playbook management, making it a reliable solution for cybersecurity automation and orchestration.

\subsubsection{Reproducibility - Open Source Offering}

To ensure the reproducibility of our results and facilitate further research and development, we have made our implementation publicly available as an open-source project. By providing access to the source code, we enable researchers, practitioners, and cybersecurity professionals to validate our findings, extend the system with new functionalities, and adapt it to their specific needs.

Beyond reproducibility, this open source initiative contributes to the broader cybersecurity community by offering a practical implementation of a CACAO-compatible KMS. As CACAO is still an emerging standard, real-world applications and collaborative development efforts are essential for its adoption. By sharing our work, we aim to accelerate the adoption of CACAO playbooks, promote best practices in playbook management, and encourage the integration of automation in cybersecurity operations.

The repository containing the source code, documentation, and installation guide is available at \href{https://github.com/Orestistsira/cacao-knowledge-base}{this link}.

\section{Discussion}

\subsection{Limitations}

The Proof-of-Concept (PoC) for the KMS was developed without user authentication, as the primary focus of this work is on ensuring efficient storage, retrieval, usage, and overall management of CACAO playbooks. This decision was made because the core research objective is to evaluate how well the system manages playbooks, rather than focusing on user identity management. Authentication mechanisms, while essential in a production environment, do not directly impact the PoC’s ability to validate the KMS functionality. Therefore, it is practical to exclude them from the current implementation.

Another limitation encountered during development was the absence of a fully developed Publish/Subscribe (Channels) model in the TAXII 2.1 specification. While TAXII provides robust mechanisms for CTI sharing, it lacks a dedicated channel-based model that would enable automated and dynamic playbook distribution. To address this gap, the Sharing Management Service was implemented, leveraging the existing Collections model to facilitate controlled playbook sharing. This service ensures that shared playbooks are properly tracked by maintaining a comprehensive sharing history in the database, preventing duplication and ensuring consistency across versions.

Moreover, the extension used for sharing CACAO playbooks as STIX Course of Action objects is not widespread enough and has not been adopted by many applications to test the sharing of playbooks under real-world conditions.

\subsection{Suggestions}

One of the points that could be improved in the CACAO standard is the inclusion of mechanisms to ensure the protection of personal data when sharing playbooks. Playbooks often contain information related to user data, such as IP addresses, usernames, or other sensitive information involved in security incident response procedures. To achieve this goal, a process could be added to pseudonymize or anonymize the sensitive data contained in playbooks before they are shared with external entities or partners. This would ensure that information related to individuals cannot be directly linked to them without the use of additional data, thus reducing the risk of privacy breaches.

Furthermore, playbooks may include one or more Data Markings, which define specific rules on how they can be shared, depending on the information they contain. However, there is no specific way in which a playbook containing sensitive information can be modified to be shareable without violating the Data Markings. This is an important issue, as in many cases sensitive information cannot simply be removed or modified without altering the functionality of the playbook.

Additionally, the CACAO standard mentions the existence of events that trigger the execution of playbooks, but does not implement any specific functionality to define or manage them. Adding such a feature would be particularly useful, especially for the system developed in this work. More specifically, the application could define various security events, such as anomaly detection or a security rule violation, which would automatically trigger one or more playbooks without the need for human intervention.

This would allow the playbooks to be automatically executed based on predefined conditions, improving real-time system response and saving valuable time when dealing with threats. This functionality would make CACAO more flexible and robust, enabling the automation of response to cyber threats with greater speed and accuracy.

A further improvement that could enhance the CACAO standard is the refinement of the playbook labeling mechanism. Currently, the standard defines labels as a playbook property, which can provide search and filtering capabilities. However, it specifies that these labels may be determined by users, organizations, or trust groups, leaving their interpretation outside the scope of the specification.

With the increasing use of playbooks for security incident response and CTI sharing, a more structured and standardized approach to labeling is necessary to improve the categorization and retrieval of playbooks. In particular, the addition of new sophisticated labels associated with CTI, such as threat types, attack types, targets, attack methods, and specific vulnerabilities, will enable easier playbook search based on specific criteria and help users identify the desired playbooks faster and more accurately.

\section{Conclusions \& Next Steps}

This work has underscored the significance of developing a KMS for CACAO playbooks, addressing the challenges encountered throughout their lifecycle. By implementing the KMS, organizations can effectively store, retrieve, execute, and share playbooks, thereby strengthening their ability to automate and orchestrate responses to cyber threats. This system not only facilitates structured playbook management but also enhances collaboration and interoperability, ultimately contributing to more efficient and coordinated cybersecurity operations.

A key factor in the success of this approach has been the implementation of the proposed KM model, which effectively meets all the requirements for a robust and practical KMS. This model establishes the necessary structure and methodologies for organizing and utilizing knowledge, ensuring systematic management of playbooks and their continuous availability for mitigating cyber threats. By providing a well-defined framework, the KM model enhances the efficiency of the system and offers a structured approach to playbook management, improving accessibility, usability, and overall effectiveness.

The paper also demonstrated that the CACAO standard offers a solid framework for playbook creation, though there are areas where improvements can be made, as previously discussed. It highlights the potential benefits of integrating modern cybersecurity tools, accelerating response processes and enhancing the resilience of organizations against evolving threats.

Future development of the system will prioritize the inclusion of authentication mechanisms and the implementation of Role-Based Access Control (RBAC). This will ensure that different users or systems have appropriate permissions when interacting with the application, based on their assigned roles. By integrating these security measures, the system will enhance access management, protecting sensitive playbooks while maintaining usability and flexibility for authorized users. 

An additional future step will be the development of a mechanism for monitoring and reporting the progress of playbook execution by SOARCA in a more refined and detailed manner. This includes real-time step tracking, detailed recording, and result reporting for auditing or debugging purposes. Additionally, the user interface will be improved to provide better visibility and management of playbook executions.

Another key enhancement will be the implementation of a Playbook Recommender System, also employing artificial intelligence. Such a system would be able to recommend specific playbook(s) given the situation at hand (e.g., specific type of incident faced or CTI information received), and even personalize search and retrieval by analyzing each user's history and data, allowing for more efficient playbook discovery. Furthermore, a Publish/Subscribe model can be introduced to improve playbook sharing within the TAXII 2.1 Server, enabling more dynamic and efficient distribution based on user preferences.

Finally, an improved Version Control feature can be developed. This will include the ability to compare version differences and insert comments or messages with each new change, ensuring better tracking and collaboration during playbook development and refinement.

\section*{Acknowledgements}

This work has received funding  from the European Union’s Digital Europe Programme (DIGITAL) under the Grant Agreement No. 101145874 (NG-SOC project) \& under the Grant Agreement No. 101128017 (CY-TRUST project), as well as from the Research and Innovation Fund of the Republic of Cyprus, under contract PRE-SEED/0823/0183. Views and opinions expressed are, however, those of the authors only and do not necessarily reflect those of the European Union or the granting authorities. Neither the European Union nor the granting authorities can be held responsible for them.






\end{document}